\documentclass[runningheads]{cl2emult}

\usepackage{makeidx}  % allows index generation
\usepackage{graphicx} % standard LaTeX graphics tool
\usepackage{subeqnar} % subnumbers individual equations
\usepackage{multicol} % used for the two-column index
\usepackage{cropmark} % cropmarks for pages without
\usepackage{math}     % placeholder for figures
\makeindex            % used for the subject index
\begin{document}

\title*{Crashes : symptoms, diagnoses and remedies}

\author{ Marcel Ausloos\inst{1}, Kristinka Ivanova\inst{2} \and Nicolas
Vandewalle\inst{3} }

\authorrunning{M. Ausloos, K. Ivanova and N. Vandewalle}

\institute{ SUPRAS \& GRASP, B5, University of Li$\grave e$ge, B-4000 Li$\grave
e$ge, Belgium \and Pennsylvania State University, University Park PA 16802, USA
\and GRASP, B5, University of Li$\grave e$ge, B-4000 Li$\grave e$ge, Belgium,}

\maketitle

\begin{abstract} A brief historical perspective is first given concerning
financial crashes, - from the 17th till the 20th century. In modern times, it
seems that log periodic oscillations are found before crashes in several
financial indices. The same is found in sand pile avalanches on Sierpinski
gaskets. A discussion pertains to the after shock period with 
illustrations from
the DAX index. The factual financial observations and the laboratory ones allow
us some conjecture on symptoms and remedies for discussing financial crashes
along econophysics lines. \end{abstract}


\section{Introduction}

Because of its magnitude, the stock market crash of Oct. 19, 1987 outshines all
downturns ever observed in the past. In one day, the Dow Jones lost 21.6~\% and
the worst decline reached 45.8~\% in Hong Kong. By comparison, with the most
famous crash of 1929, the crash was spread over 2 days: The Dow Jones sank
12.8~\% on October 28 and 11.7~\% the following day. This shows that a stock
market index decline does not necessarily lead to a crash in one 
single day. The
decline can be slow and last several days or even several months in 
what would be
called not so drastic cases, but a long duration bear market. Notice 
that markets
had been using {\it breakers} or {\it periods of trading halts and 
limitations of
daily variations} in order to avoid anomalous drops. However, Lauterbach and
Ben-Zion \cite{lb1993} found that trading halts and price limits had 
no impact on
the overall decline of October 1987.

Such financial factual observations should be turned into 
quantitative measures.
That is why we looked at the similarities between index evolution in 87 and 97,
as early as spring 97 \cite{dupuis,viz}. Previous historically famous crashes
served as a thinking basis, - see Sect. 2. In Sect. 3 it is recalled that
log-periodic corrections are sometimes superimposed on an index trend {\it
before} a crash. In Sect. 3 the types of fluctuations in a {\it post}-crash
period are also touched upon. Sorting out distinct behaviors should 
be made {\it
a priori}, as here below for the DAX.

Fractal systems provide a good modelling of a wide class of media in particular
those containing a hierarchical structure, like in financial markets
\cite{hierarchy}. We conjecture that the Bak-Tang-Wiesenfeld (BTW) model
\cite{btw} of a sandpile, generalized for fractal bases contains several
ingredients which could be translated {\it mutatis mutandis} into a model for
stock market index evolution because such a model contains log-periodic
oscillations as found before crashes. Results on the BTW model \cite{btw} on
prefractal Regular Sierpinski Carpets (RSC) having various fractal 
dimensions and
connectivity are reported in Appendix.  Since {\it log-periodic corrections}
characterizing features in the distribution of sand avalanches depend on the
ramification of the RSC, this allows us to propose reflection lines on remedies
against crashes in Sect. 4.

\section{Some Historical Notes}

Although there might have been financial crises in previous times and locations
all over the world, the most famous one in {\it recent times} is the 
Tulipomania
and subsequent crash which occurred in the 17-th century in Holland
\cite{webtulip}. Everything started in 1559 when the first tulip bulb (TB) was
brought to Holland. The flower was considered so rare that 
speculation ensued and
the flower became wildly overvalued : in 1635, 1 TB was worth 4 tons 
of wheat + 4
oxen + 8 tons rye + 8 pigs + one bed + 12 sheep + clothes + 2 wine 
casks + 4 tons
beer + 2 tons butter + 1000 pounds cheese + 1 silver drinking cup. In 
1637, 1 TB
was worth 550 NLG, i.e. a 117~\% return/year. However within 1637, in a 6 week
time the price of 1 TB went down 90~\%. Nowadays a black tulip bulb costs about
0.5 EUR.

A set of similar financial crises is that made of the Compagnie du Mississipi
\cite{weblaw} and that of the South Sea Company 
\cite{websouthsea1,websouthsea2}
bubbles. In 1715, John Law had persuaded Philippe, Regent of France, 
to consider
a banking scheme that promised to improve the financial condition of 
the kingdom.
In theory a private affair, the system was linked from the beginning with
liquidating the national debt. When the monopoly of the Louisiana trade was
surrendered in 1717, Law created a trading company known as the {\it Compagnie
d'Occident} (or {\it Compagnie du Mississipi}) linked to the bank and in which
government bills were accepted for the purchase of shares, Law 
gaining a monopoly
on all French overseas trade. The result was a huge wave of speculation as the
value of a share went from its initial 500 {\it livre} value to 18 
000 $livres$.
When the paper money was presented at the bank in exchange for gold, which was
unavailable, panic ensued, and shares felt by a factor of 2 in a 
matter of days.

In England, the Whigs invented the Bank of England in 1694. The South 
Sea Company
(SSC) was formed in 1711 by the Tory government to trade with Spanish America,
and to offset the financial support which the Bank of England had provided for
Whig governments. The Tories had in mind to establish a system like 
the Compagnie
du Mississipi Monopoly \cite{weblaw}, using trading privileges and monopolies
granted to Britain after the Treaty of Utrecht. In 1720 a Parliament bill was
passed enabling persons to whom the government owed portions of the 
national debt
to exchange their claims for shares in SSC $stocks$. On March 1 the SSC stocks
were valued GBP 175. On June 1, the shares were valued 500 and more 
than 1 000 in
August 1720. Speculators took advantage of investors to obtain 
subscriptions for
patently impossible projects. By September 1720 however, the market had
collapsed, and by December SSC shares were down to 124, dragging 
other, including
government, stocks with them. Many investors were ruined including I. Newton
\cite{Newton}.

In the years from 1925 to 1929 it was almost a craze to play the market
\cite{wallstreetcrash29}. One could go to a broker and purchase stock 
on margin.
That allowed the speculation bubble to grow unchecked. Many $mini$ crashes and
subsequent recoveries began as early as Monday March 25, 1929. The 
summer of 1929
was not too bad. However on Sept. 3, a bear market became firmly 
established, and
on Thursday Oct. 24, 1929 a crash occurred.  In fact such crashes, 
not mentioning
the uncontrolled buying frenzy on IPOs stocks at the end of the 1990's in
companies for which owners do not have a coherent business plan, look 
all similar
: euphoria and speculation.

Other recent summers (rather than octobers!) are in the memory of 
investors, e.g.
the CAC 40 dropped every summer between 1990 and 1998, except for 
1993 (Table 1).
There were 11 declines on the S\&P 500 since 1925. One of these were horrendous
($ca.$ 43~\% between Jan. 73 and Sept. 74). In so doing it can be 
emphasized that
crashes can be very abrupt but a market drop of the same, and even bigger
importance can also occur.

\begin{table}[ht] \centering \caption{The drop in CAC 40 values in the 1990's.
The ''date'' is at the start of the ''drop'', given in \% and $\Delta t$ is the
time interval over which the drop occurred} \renewcommand{\arraystretch}{1.4}
\setlength\tabcolsep{5pt} \begin{tabular}{ccccccccc} \hline Year & 1990& 1991&
1992& 1994& 1995& 1996& 1997& 1998\\ date & Aug.02& Aug.18& Jun.02& Aug.30&
Aug.21& Jun. 30& Jul.01& Jul.20\\ drop &  20 & 8 & 12 & 12 & 11 & 9 & 
10 & 20 \\
$\Delta t$ & 3 w &3 d &2 m &1 m &1 m &1 m &2 m &1 m\\ \hline \end{tabular}
\label{Tab1} \end{table}

\section{Empirical Universality and Symptoms}

Bates \cite{Bates1991} studied transactions prices of S\&P 500 futures options
{\it a posteriori} over 1985-1987 to find out expectations/precursors 
of a shock.
He discovered that $out-of-money$ $puts$ were unusually expensive to
$out-of-money$ $calls$. Then, the use of a jump-diffusion model for 
daily options
prices of 1987 leaded to the conclusion that an expected negative jump was
predictable a year prior to the crash. Other techniques exist, like 
those looking
at the probability distribution function of returns \cite{pdf,Lillo}. The main
difference with our studies \cite{viz,phasetr,how}, and that of others
\cite{SJB96,FF96,FF98,BC98,GY98,Can00}, comes from the {\it a priori} 
analysis of
the 1990-1997 scenario, trying to find out if a break point in an index series
becomes more and more probable, $how ?$, and may be $why?$. These ideas are
controversial \cite{Laloux98,JLS98}.

\subsection{Before: diagnoses}

The application of statistical physics ideas to the forecasting of stock market
behavior and crashes has been proposed earlier 
\cite{phasetr,SJB96,FF96}. It was
proposed that an economic index $y(t)$ could increase as a complex power law
\cite{SJB96}, i.e.

\begin{equation} y = A + B {\left( {t_c-t \over t_c} \right)}^{-m} \left[ 1 + C
\sin\left( \omega \ln{\left( {t_c-t \over t_c} \right)} + \phi \right) \right]
\hskip 0.5cm {\mbox{for}} \hskip 0.3cm t < t_c \end{equation}

\noindent where $t_c$ is the crash-time(day) or rupture point, $A$, $B$, $m$,
$C$, $\omega$ and $\phi$ are free parameters, while the period of the
oscillations converges to the rupture point at $t=t_c$. This law is similar to
that of critical points at so-called second order phase transitions
\cite{StanleyPTbook} but with a complex exponent $m + i \omega$, and 
generalizes
the scaleless situation  discrete scale invariance cases \cite{DSphysrep,JSL}.
 From the stock market point of view, the equation has been derived Canessa
\cite{CanessaRNG} along renormalization group lines.

The S\&P500 data \cite{how} for the period preceding the 1987 October 
crash were
already fitted using Eq.(1). It has been stressed that a nonlinear 
parameter fit
does not easily lead to robust values against small data perturbations
\cite{macdo}, the more so when there are seven parameters. Various 
values of $m$,
including negative ones, were in fact reported in the literature for various
indices and events \cite{phasetr,how,buda,jura,ladek}, and are summarized in
Table 2.

\begin{table}[ht] \centering \caption{Values of the coefficients in 
Eq. (1) that
result from fitting different financial indices to Eq. (1)}
\renewcommand{\arraystretch}{1.4} \setlength\tabcolsep{5pt}
\begin{tabular}{clccccc} \hline Period & Index& $m$ &$A$ & $B$ & $t_c^{div}$ &
$R$\\ \hline 80-87&Dow&0&-499.4$\pm$16.1&-532.9$\pm$5.6&87.85$\pm$0.02&0.951\\
80-87&Dow&1/3&-526.6$\pm$20.8&614.7$\pm$8.6&88.22$\pm$0.03&0.956\\
80-87&Dow&1/2&-5.7$\pm$15.2&257.8$\pm$4.1&88.46$\pm$0.04&0.956\\
80-87&S\&P500&0&-57.4$\pm$2.5&-68.9$\pm$0.9&87.89$\pm$0.03&0.947\\
80-87&S\&P500&1/3&-80.3$\pm$3.7&88.2$\pm$1.6&88.45$\pm$0.04&0.949\\
80-87&S\&P500&1/2&-11.6$\pm$2.8&38.8$\pm$0.8&88.78$\pm$0.05&0.949\\
80-87&FTSE&0&-563.5$\pm$31.9&-512.9$\pm$9.8&87.85$\pm$0.03&0.960\\
80-87&FTSE&1/3&-449.9$\pm$41.4&549.9$\pm$15.1&88.21$\pm$0.05&0.958\\
80-87&FTSE&1/2&59.1$\pm$31.5&222.3$\pm$7.1&88.41$\pm$0.06&0.956\\
90-97&Dow&0&-1919.6$\pm$38&-1762$\pm$13.4&97.92$\pm$0.02&0.978\\
90-97&Dow&1/3&-2100.4$\pm$49&2081.8$\pm$20.3&98.39$\pm$0.03&0.982\\
90-97&Dow&1/2&-360.1$\pm$35.8&882$\pm$9.7&98.68$\pm$0.04&0.982\\
90-97&S\&P500&0&-141.5$\pm$4.4&-187$\pm$1.5&97.90$\pm$0.02&0.974\\
90-97&S\&P500&1/3&-161.4$\pm$6.1&221.3$\pm$2.5&98.38$\pm$0.03&0.976\\
90-97&S\&P500&1/2&23.2$\pm$4.5&93.9$\pm$1.2&98.67$\pm$0.04&0.976\\
90-97&FTSE&0&-499.1$\pm$46.9&-1109.9$\pm$19&98.44$\pm$0.04&0.951\\
90-97&FTSE&1/3&-1310$\pm$86&1633.3$\pm$40.9&99.51$\pm$0.08&0.948\\
90-97&FTSE&1/2&-189.8$\pm$66.3&770.6$\pm$22.3&00.10$\pm$0.10&0.948\\ \hline
\end{tabular} \label{Tab2} \end{table}

It would be nice stipulating that $m$ could be $universal$ by analogy 
with second
order phase transitions, at least for the presently studied crashes, seemingly
falling into {\it financially similar classes}, even though this is surely an
unrealistic dream. A behavior which we considered was the logarithmic 
divergence

\begin{equation} y = A + B \ln{\left( {t_c-t \over t_c} \right)} \left[ 1 + C
\sin\left( \omega \ln{\left( {t_c-t \over t_c} \right)} + \phi \right) \right]
\hskip 0.5cm {\mbox{for}} \quad t < t_c . \end{equation}

As in critical point data analysis the optimum test consists in separating the
most diverging term from the others, after having eliminated the so called {\it
mean field trend} and searching for the correction to scaling \cite{brezin}. In
fact, the fit can be made in two steps : (i) one looks for a $t_c^{div}$, i.e.
for the logarithmic divergence; then (ii) for $t_c^{osc}$ for the oscillation
convergence \cite{how}.

Due to the log-periodicity in Eq.(2), the relation

\begin{equation} t_c^{osc} = {t_n - {t_{n+1} \over \lambda} \over 1 - {1 \over
\lambda}} \end{equation}

\noindent holds true where $\lambda=\exp{(\omega/2\pi)}$ and $t_n$, 
$t_{n+1}$ are
successive maxima or minima days (Table 3). The results \cite{how} readily show
that the examined stock market indices well follow a logarithmic law 
divergence.
It should be noted that $t_c^{osc}$ and $t_c^{div}$ are extremal 
dates since the
index should necessarily fall down before it reaches infinity. 
Moreover for both
1980-87 and 1990-97 period cases, it is found that the value of $\lambda$ seems
to be almost constant, (Table 3), corresponding to $\omega \simeq 6$. 
An analysis
along similar lines of thought, though emphasizing the no-divergence was
discussed for the Nikkei \cite{sornettenikkei,stauijtaf} and NASDAQ April 2000
crash \cite{sornetteNASDAQ}.

\begin{table}[ht] \centering \caption{The $\lambda$ and $t_c^{osc}$ values
obtained for three indices following the methodology explained in the text. The
real rupture point of Oct. 19, 1987 is $t_c$=87.79, and that of Oct. 24, 97 is
$t_c$=97.81} \renewcommand{\arraystretch}{1.4} \setlength\tabcolsep{1pt}
\begin{tabular}{ccccccc} \hline Period& &80-87& & &90-97& \\ \hline 
Index	&
Dow &S\&P500&FTSE& Dow &S\&P500&FTSE \\ $\lambda$ & 2.382$\pm$0.123&
2.528$\pm$0.127& 2.365$\pm$0.137& 2.278$\pm$0.045& 2.549$\pm$0.163&
2.3745$\pm$0.054 \\ $t_c^{osc}$ &87.91$\pm$0.10 &87.88$\pm$0.07 &87.87$\pm$0.10
&97.89$\pm$0.06 &97.85$\pm$0.08 &97.85$\pm$0.05\\ \hline \end{tabular}
\label{Tab3} \end{table}

\subsection{During}

First consider that a crash can occur under four different conditions, and be
listed in four categories, i.e. PMP, PMM, MMP, MMM, where M and P indicate an
index variation from one day to another. The middle variation represents the
crash amplitude. This allows for $mini$ and $maxi$ crashes. For this report
consider the DAX variations between Oct. 1, 1959 and Dec. 30, 1998. In Figs.
1(a-d) we show the DAX partial distribution of fluctuations (pdf) 
resulting from
distinguishing such categories. The pdf's have fat tails far from a Gaussian
distribution and scale as a power law with exponent $\mu$, $P(g(i))\sim
g(i)^{-\mu}$, where $g(i)=log(y(i+1)/y(i))$ and $y(i)$ denotes the signal.
Approximate values of the $\mu$ exponent are given in Table 4. The nine most
drastic crashes in each category are shown in Table 5 according to the value of
$g(i)=log(y(i+1)) - log(y(i))$, together with the corresponding day. 
Notice that
the case studied by Lillo et al. \cite{Lillo}, occurring on Aug. 31, 
1998 is not
included among these 36 crashes.

\begin{figure} \centering 
\includegraphics[width=.48\textwidth]{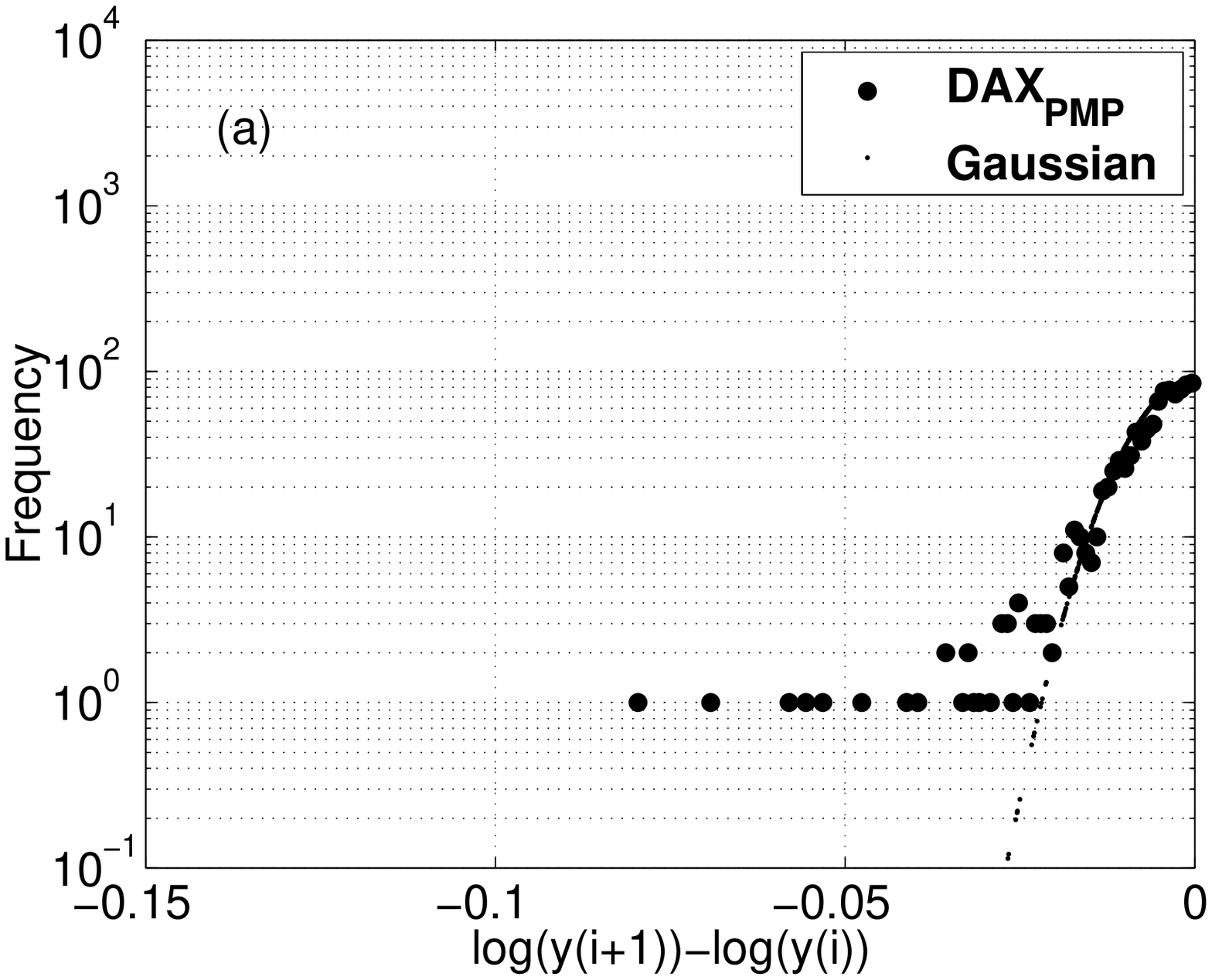} \hfill
\includegraphics[width=.48\textwidth]{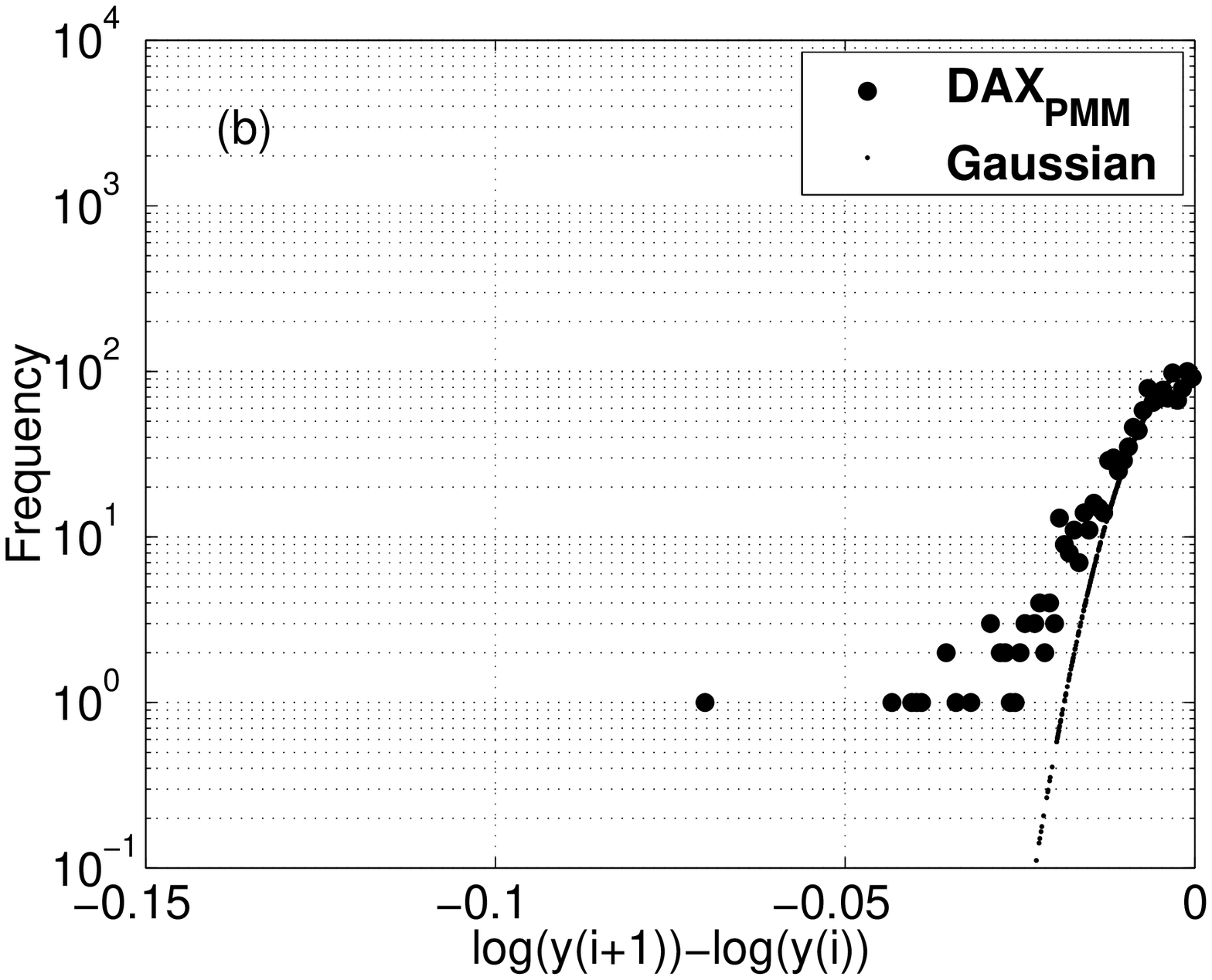} \vfill
\includegraphics[width=.48\textwidth]{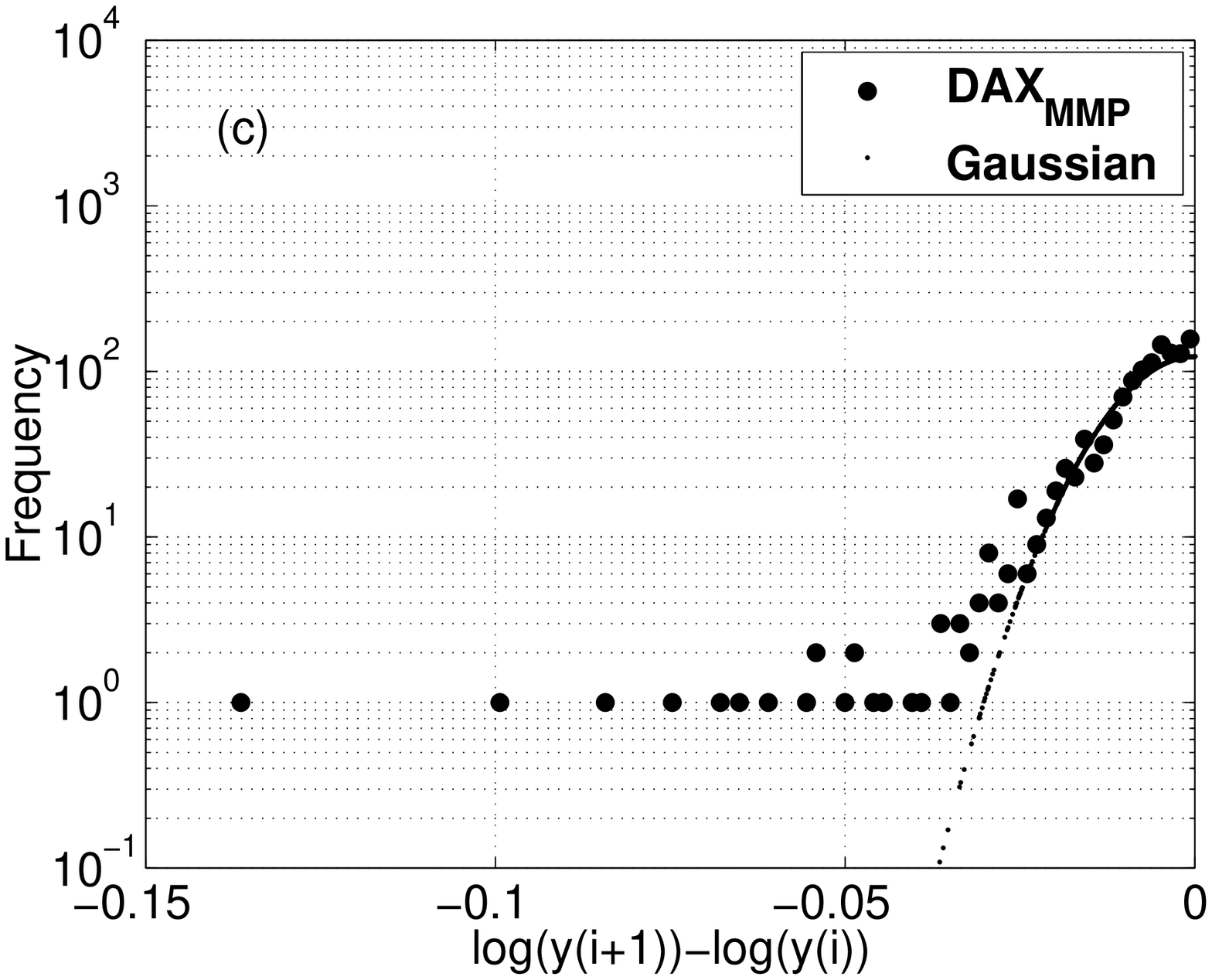} \hfill
\includegraphics[width=.48\textwidth]{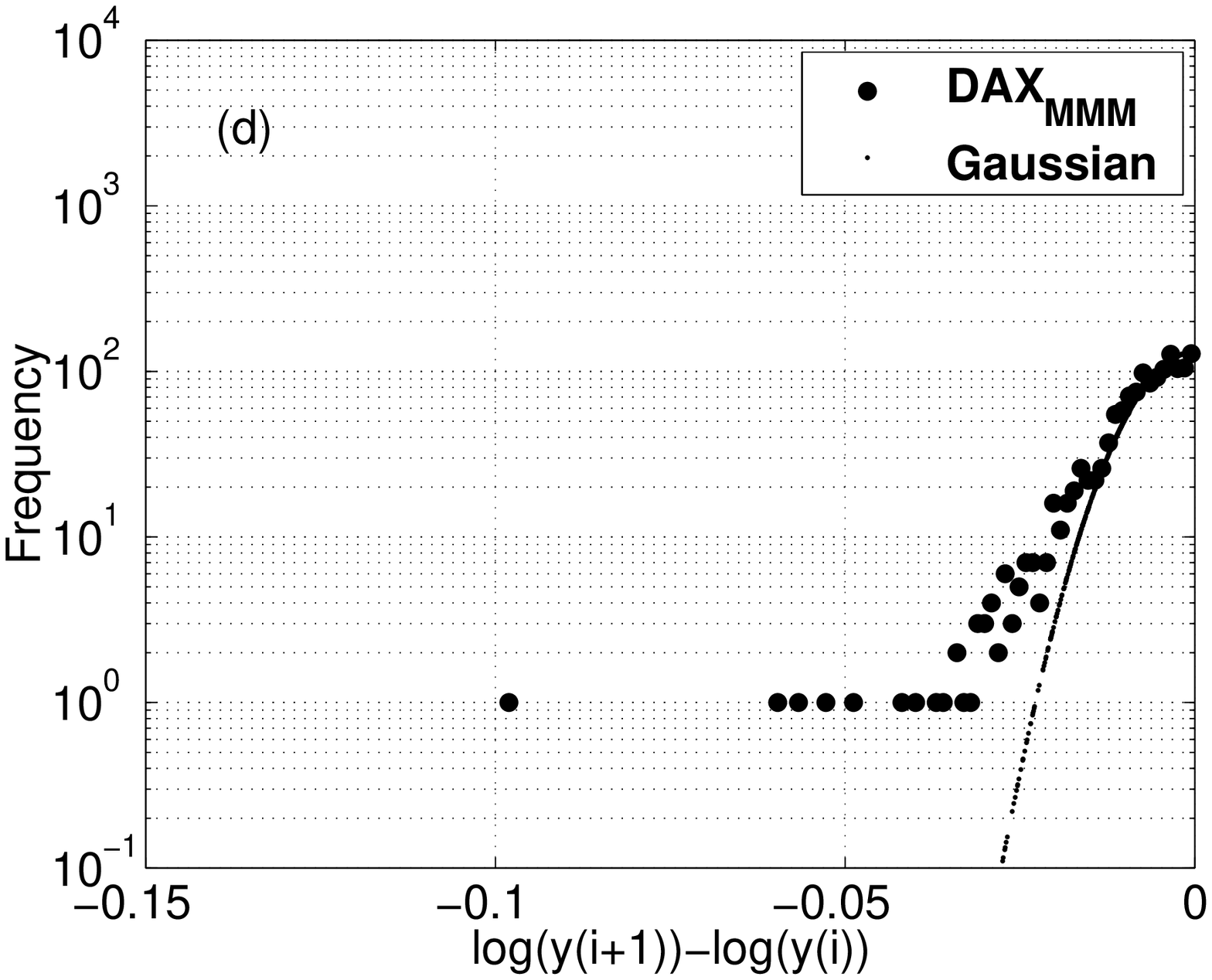} \caption{The 
distribution of the
fluctuations of {\bf (a)} PMP, {\bf (b)} PMM, {\bf (c)} MMP, and {\bf (d)} MMM,
as compared to the Gaussian distribution for the DAX between Oct. 1, 
1959 to Dec.
30, 1998. {\it Dotted lines} correspond to a Gaussian fit to the 
central part of
the distributions} \label{eps1} \end{figure}

\begin{table}[ht] \centering \caption{The $\mu$ exponent of the pdf 
tail's power
law dependence, the mean spectral exponent $\beta$ and the corresponding mean
fractal dimension $D$ for the DAX 600 day long recovery signals after the crash
for each crash category, i.e. PMP, PMM, MMP and MMM of the DAX between Oct. 01,
1959 and Dec. 30, 1998.} \renewcommand{\arraystretch}{1.4}
\setlength\tabcolsep{5pt} \begin{tabular}{cccc} \hline case & $  \mu$ 
& $<\beta>$
& $<D>$\\ \hline PMP& 2.76$\pm$0.12& 1.70$\pm$0.38 & 1.65$\pm$0.38\\
PMM&2.76$\pm$0.19&1.79$\pm$0.30&1.60$\pm$0.30\\
MMP&2.85$\pm$0.17&1.60$\pm$0.36&1.70$\pm$0.36\\
MMM&2.83$\pm$0.23&1.68$\pm$0.35&1.66$\pm$0.35\\

\hline \end{tabular} \label{Tab4} \end{table}

\begin{table}[ht] \centering \caption{The nine strongest crashes in each PMP,
PMM, MMP and MMM category listed in decreasing order strength measured by the
value of $g(i)=log(y(i+1)) - log(y(i))$ for DAX between Oct. 1, 1959 
and Dec. 30,
1998} \renewcommand{\arraystretch}{1.4} \setlength\tabcolsep{3pt}
\begin{tabular}{ccccccccc} \hline
&\multicolumn{2}{c}{PMP}&\multicolumn{2}{c}{PMM}&\multicolumn{2}{c}{MMP}
&\multicolumn{2}{c}{MMM}\\ \hline &\multicolumn{2}{c}{\# of cases =
960}&\multicolumn{2}{c}{\# of cases = 1247}&\multicolumn{2}{c}{\# of cases =
1247}&\multicolumn{2}{c}{\# of cases = 1360}\\ \hline &date & $g(i)$ & date &
$g(i)$ &date & $g(i)$ &date & $g(i)$ \\ \hline 1.&Oct 26, 87&-0.080&Oct 28,
87&-0.070&Oct 16, 89&-0.137& Oct 19, 87&-0.099\\ 2.&Oct 22, 87&-0.069&Oct 27,
97&-0.043&Aug 19, 91&-0.099& Sep 10, 98&-0.060\\ 3.&Jan 04, 88&-0.058&Aug 22,
97&-0.040&Oct 28, 97&-0.084& Oct 01, 98&-0.057\\ 4.&Mar 06, 61&-0.056&Sep 17,
98&-0.040&May 29, 62&-0.075& Nov 09, 87&-0.053\\ 5.&Jul 07, 86&-0.053&Dec 28,
87&-0.039&Nov 10, 87&-0.068& Dec 01, 98&-0.049\\ 6.&Oct 23, 97&-0.048&Jan 11,
88&-0.036&Oct 02, 98&-0.065& May 28, 62&-0.042\\ 7.&Dec 06, 96&-0.041&Oct 22,
62&-0.035&Aug 21, 98&-0.061& Jan 14, 91&-0.040\\ 8.&Apr 01, 97&-0.040&Nov 22,
73&-0.034&Aug 06, 90&-0.056& Oct 05, 92&-0.037\\ 9.&Jan 21, 74&-0.036&Aug 17,
62&-0.032&Mar 13, 74&-0.055& Aug 17, 90&-0.036\\ \hline \end{tabular}
\label{Tab5} \end{table}

\subsection{After}

To study the index evolution after the crash we construct an evolution signal
(Fig. 2 (a-d)) that is the difference between the DAX value signal at each day
$y(i)$ and the DAX value at the crash day $y_0$ for the 36 cases of interest
reported in Table 5. For most of the crashes, i.e. all crashes that 
occur before
Aug. 06, 1996, the evolution signal is 600 day long. However, for crashes that
occur after Aug. 6, 96, e.g. less than 600 days before the last day of this
study, the evolution signal is shorter, as for example the Oct. 01, 98 MMM case
in Fig. 2d. There are 3 $short$ cases in each category.

\begin{figure} \centering 
\includegraphics[width=.48\textwidth]{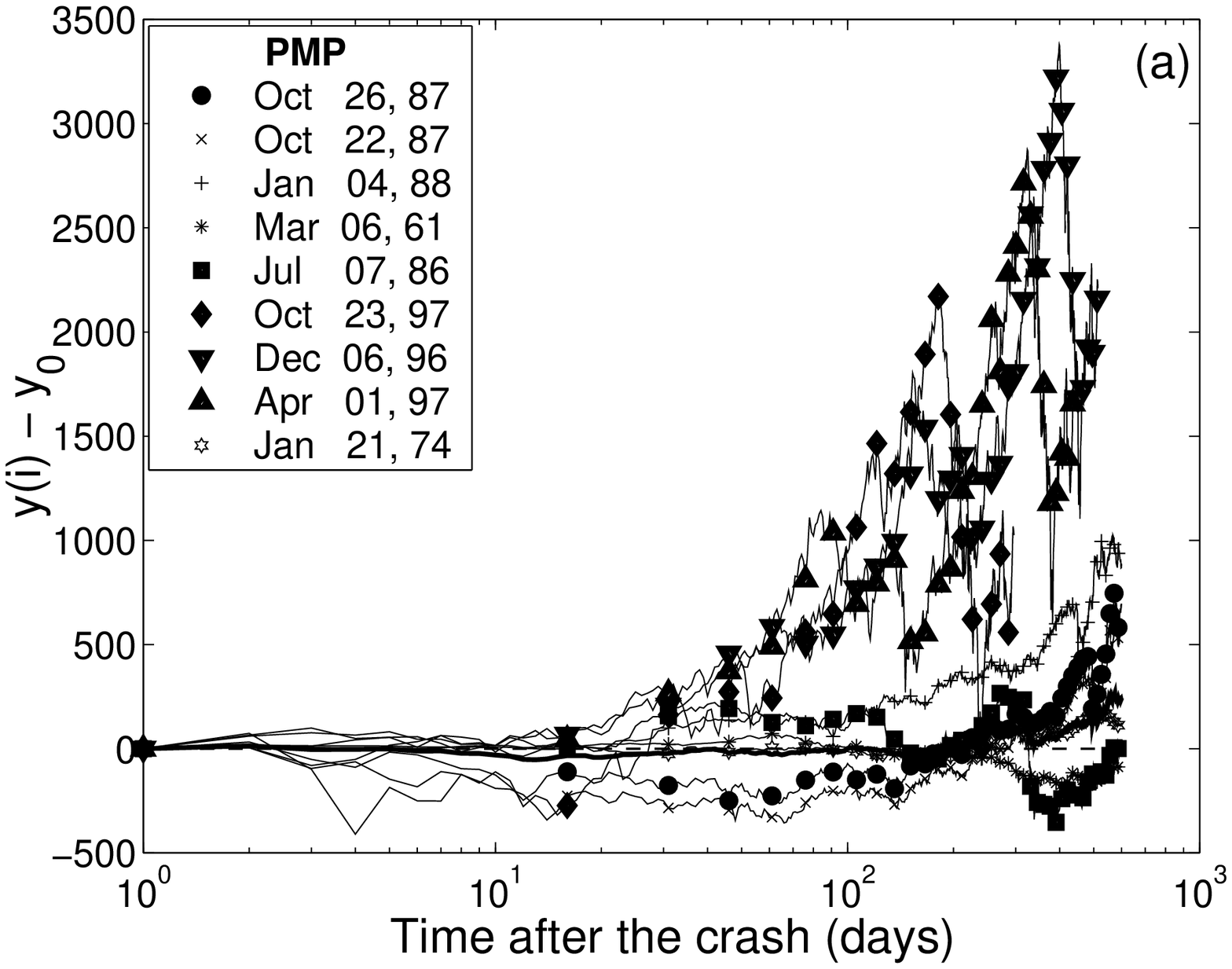} \hfill
\includegraphics[width=.48\textwidth]{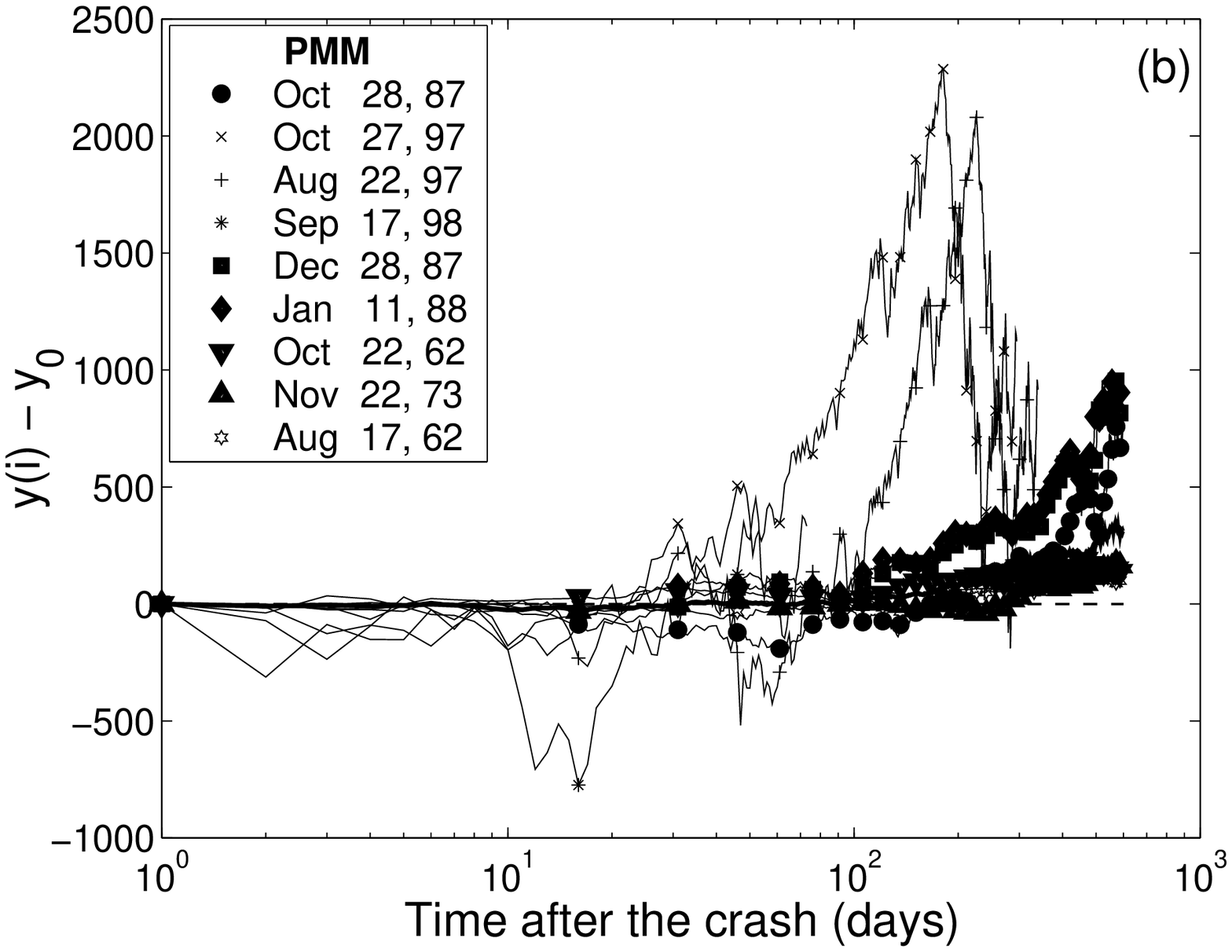} \vfill
\includegraphics[width=.48\textwidth]{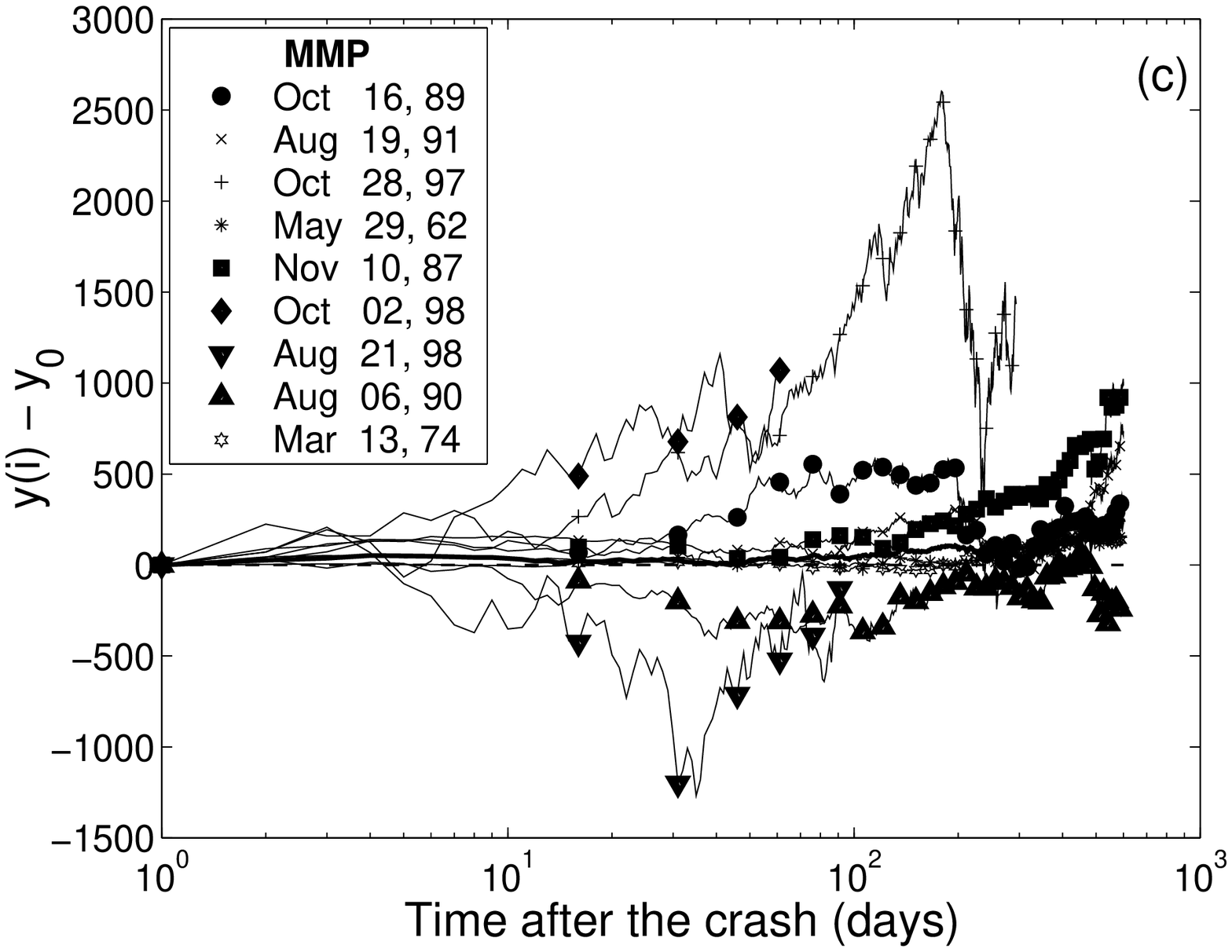} \hfill
\includegraphics[width=.48\textwidth]{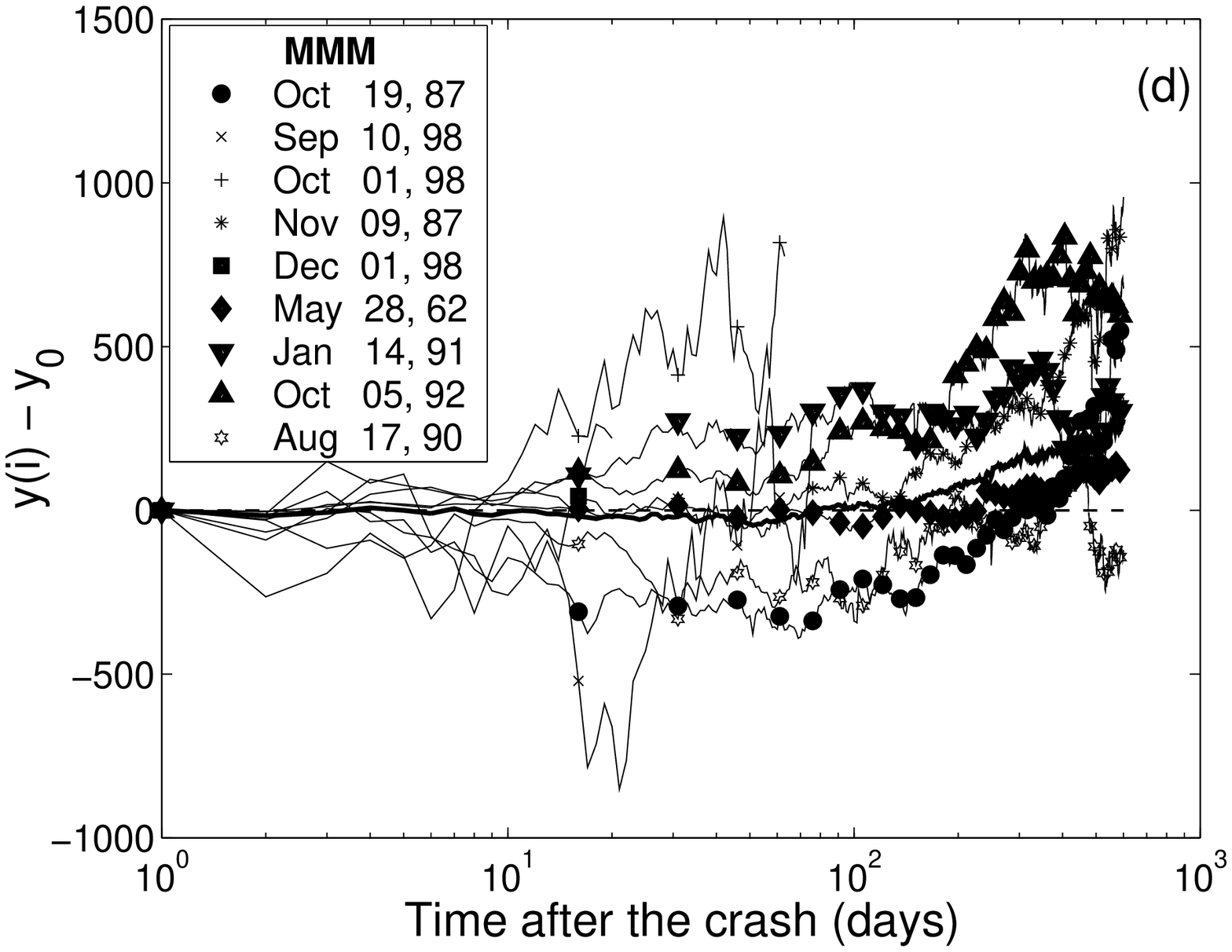} \caption{The evolution of the
nine strongest DAX crashes in {\bf (a)} PMP, {\bf (b)} PMM, {\bf (c)} MMP, and
{\bf (d)} MMM categories as listed in Table 5; $y_0$ denotes the index value at
the ''origin of recovery'', i.e. the value of the signal at the closure time of
the crash day. The {\it thick solid line} corresponds to the average 
evolution of
the recovery signal in each category} \label{eps2} \end{figure}

Notice that recovery can be slow. The PMP and PMM cases need of the order of
thirty days before having a positive $y(i)$ - $y_0$ value. The 
situation is more
complicated for the MMP and MMM cases. To see if some periodic 
fluctuation occurs
after the crash, the power spectrum of the DAX has been studied for the 600 day
long signals, i.e. for 24 cases. The power spectrum corresponding to the major
crash in each category is given in Figs. 3 (a-d). Note the high-frequency
log-periodic oscillation regime of the power spectrum for the 
strongest MMM case
that occurs on Oct. 19, 1987 on Fig. 3 d.

\begin{figure} \centering 
\includegraphics[width=.49\textwidth]{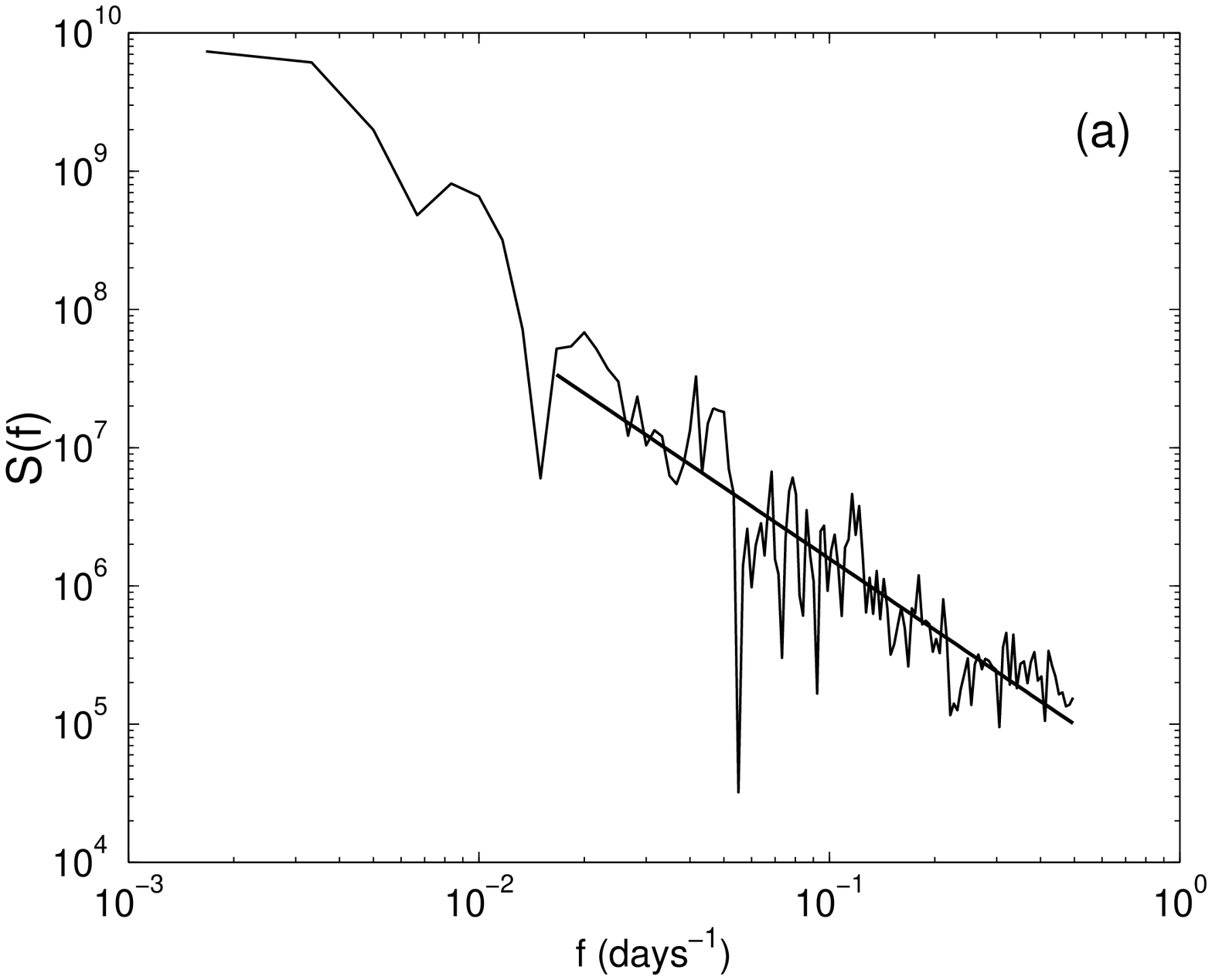} \hfill
\includegraphics[width=.49\textwidth]{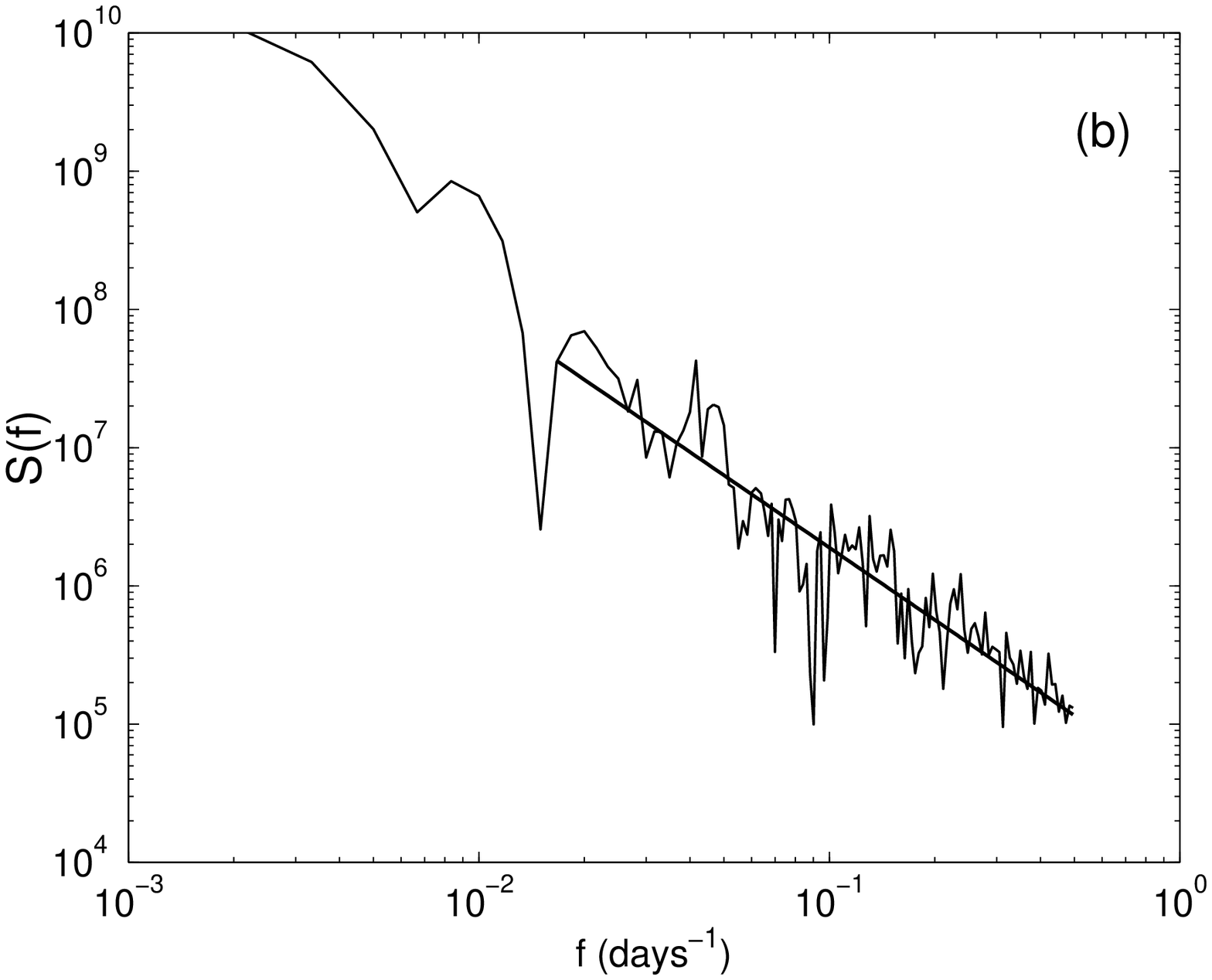} \vfill
\includegraphics[width=.49\textwidth]{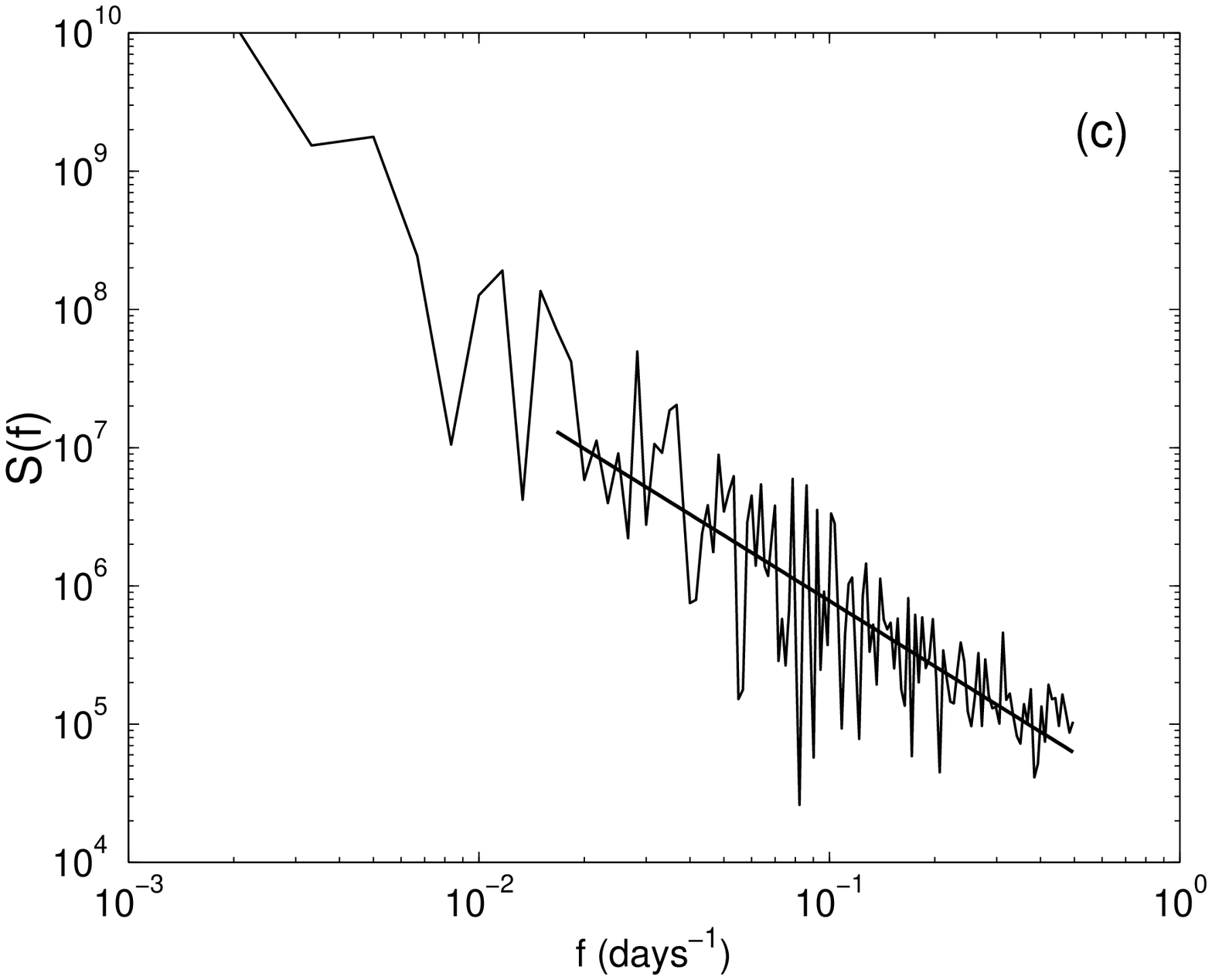} \hfill
\includegraphics[width=.49\textwidth]{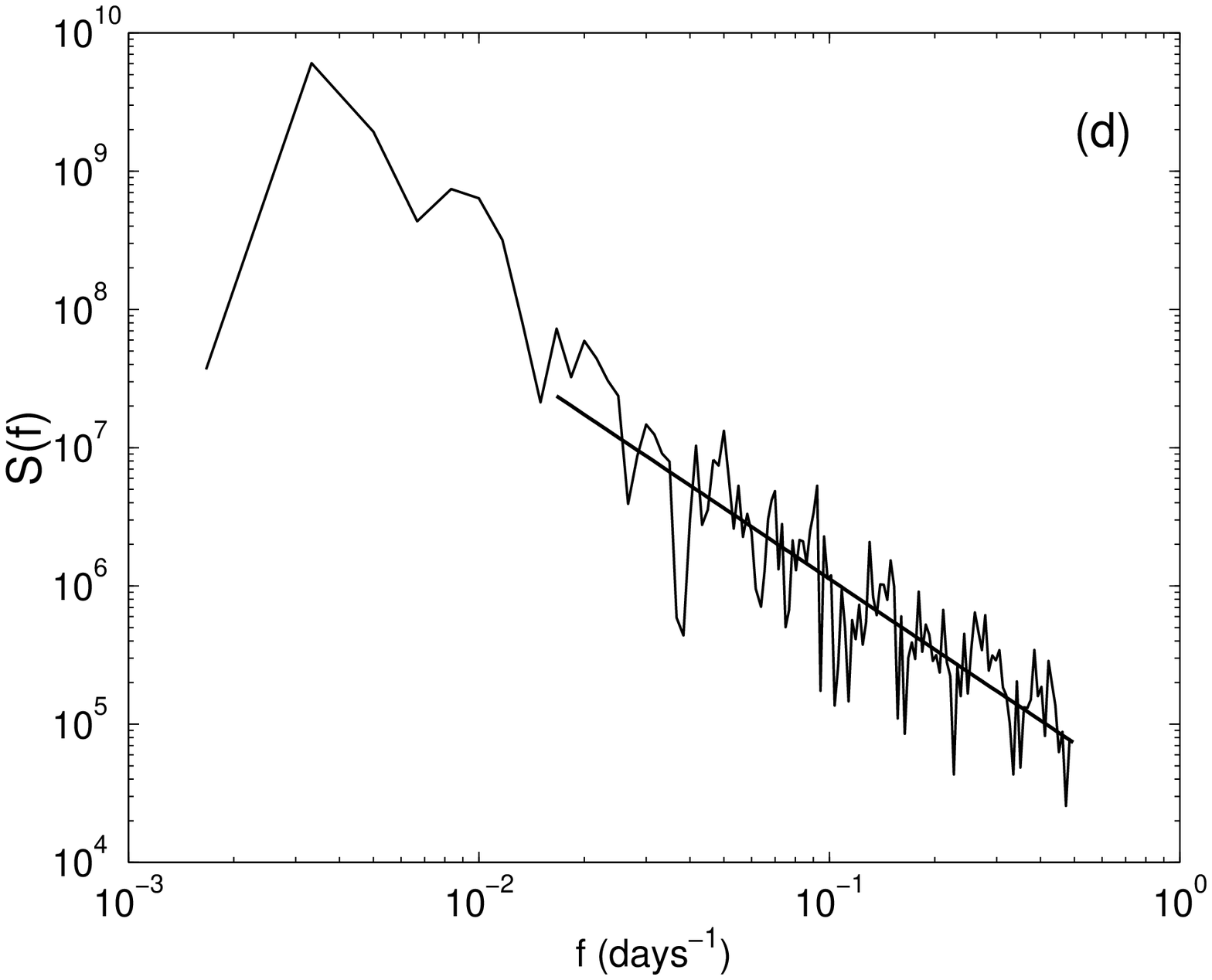} \caption{The power spectrum of
the 600 day DAX index evolution signal corresponding to the major crash in each
category {\bf (a)} PMP : Oct. 26, 1987, {\bf (b)} PMM : Oct. 28, 
1987, {\bf (c)}
MMP : Oct. 16, 1989, and {\bf (d)} MMM : Oct. 19, 1987. } \label{eps3}
\end{figure}

To estimate the behavior of the DAX index evolution signal post PMP, PMM, MMP,
MMM crashes it is of interest to relate each spectral exponent $\beta$ to the
fractal dimension $D$ of the signal through \cite{schroeder}

\begin{equation} D = E + \frac{3 - \beta}{2}, \end{equation}

\noindent
where $E$ is the Euclidian dimension. The values of the averaged $\beta$ and
averaged fractal dimension $D$ are reported in Table 4.

\section{Predictability and remedies for a conclusion}

Let us assume for the following arguments that one can discuss stock market
crashes in terms of physical model considerations. Moreover, let us admit that
signals can be treated as above, in terms of power laws, and 
oscillations. In so
doing we use the framework which has been useful in analyzing the avalanche
problem for sand piles in the Appendix. Let us wonder whether these
considerations, and analogies can suggest remedies in order to control or even
avoid crashes. It is easily argued that remedies can be either self-remedies or
due to external fields. At thermodynamic phase transitions, impurities, or
external fields can shift the critical temperature, and reduce the 
divergence of
thermodynamic properties. Let us disregard here the case of external field,
though several authors might consider that in some economies such fields are
relevant, or more necessary than self-corrections.

Several variables, or parameters, are to be considered : (i) the time scale, or
frequency $\omega$, (ii) the amplitudes of the signal $A_i$, (iii) the
dimensionality of the system, (iv) the connectivity $\lambda$ of the 
lattice. The
amplitude is related to the "amount of sand" or volume exchanged during
transactions, while the connectivity is related to gradient of trades, which is
somewhat similar to the sand "angle of repose".

First it is absolutely clear that if the sand flux is large, an 
avalanche will be
very likely, and disordered. This can be seen to be analogous to the 
retention of
orders on a market, and to the effect of breakers \cite{grassia}. Contrary to
reducing, stopping the avalanche effect, breakers are in fact accelerating the
process. One remedy is therefore to reduce the amount of sand, i.e. 
the number of
orders should decrease with time, and some delay should be imposed between
orders. This is similar to changing the angle of repose of the materials.

Another {\it new remedy} is hereby introduced, the connectivity. It 
has been seen
in the Appendix that when the connectivity increases the avalanche distribution
is more spread out, the log-periodicity feature is not so pronounced. 
Therefore
it seems relevant that the number of actors on the market be increased at crash
time, together with the decrease in exchanged volume. Furthermore the
connectivity is a key ingredient in the spreading of information on a market
\cite{huangkoln} and also leads to consider the effect of interacting 
agents and
herding models \cite{kaizoji,hulst}. Notice that this effect is entirely
contained in the oscillations which therefore smoothen the rate of 
divergence. It
is of interest to observe that the connectivity, in the stock market, is a {\it
rather small number}, ca. 6.0. To expect that the connectivity is a constant
whatever the hierarchical \linebreak stage is of course a utopia, but this can be 
taken as a
first approximation.

Another consideration pertains to the question whether a model and its solution
can be implemented, and if so if any crash can be avoided. Two comments are in
order. First, one has to bear with statistical physics that as long 
as hypotheses
are fulfilled the class of transition is defined, and therefore a crash will
occur if the conditions are fulfilled. Next economic and speculative
considerations will always exist. Therefore crashes will always exist
\cite{dean}, - except maybe under conditions/remedies outlined in the preceding
section.

Whether
there is an influence of a known theoretical model on a financial 
event like a crash is exactly
similar to wondering whether the equilibrium market hypothesis holds
true.

\vskip 0.2cm {\noindent \large \bf Acknowledgements} \vskip 0.2cm

We are very grateful to the organizers of the Symposium for their 
invitation and
to the Symposium sponsors for financial support. NV acknowledges a grant from
FNRS/LOTTO which allowed us to perform specific numerical work. KI acknowledges
partial support from Battelle 327421-A-N4 grant. MA had fruitful 
discussions with
J. Ph. Bouchaud, Th. Lux, P.Moussa, A. Pekalski, H.E. Stanley, and D. Stauffer.
NV had fruitful discussions with E.Clement, S.Galluccio, S.Galam, 
R.N. Mantegna,
J.Rajchenbach, and S.Zapperi.

\newpage 
{\noindent \large \bf Appendix}

\vspace*{0.3cm}

The Bak, Tang and Wiesenfeld (BTW) \cite{btw} accumulation-dissipation process
model on  regular lattices was extended to a sand pile version 
\cite{bak2,zhang}.
Recently, the BTW process was studied on a Sierpinski gasket of 
fractal dimension
$D_f=\frac{\ln{3}}{\ln{2}}$ \cite{gasket,luc}. It has been shown that the
avalanche dynamics is characterized by a power law with a complex scaling
exponent $\tau + i \omega$ with $\omega = \frac{2\pi}{\ln{2}}$. This was
understood as the result of the underlying Discrete Scale Invariance 
(DSI) of the
gasket, i.e. the lattice is translation invariant on a log-scale
\cite{DSphysrep}. It is possible to extend the study of the BTW model on other
(prefractal) Regular Sierpinski Carpets (RSC) by varying both the fractal
dimension $D_f$ as well as the connectivity of the lattice. In so doing we have
observed apparently connectivity-based corrections to power law scaling.

\begin{figure} \centering \includegraphics[width=.9\textwidth]{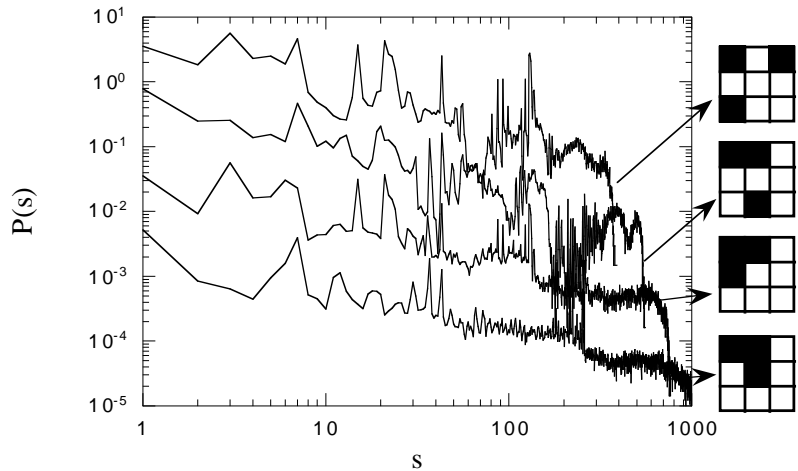}
\caption{The size distribution of avalanches $P(s)$ for the BTW model 
on fractal
RSC lattices having the same fractal dimension $D_f = \frac{\ln{6}}{\ln{3}} + i
\frac{2\pi}{\ln{3}}$. The generators of the Sierpinski carpets are indicated}
\label{eps4} \end{figure}

Four different RSC of generation $n=2$ are illustrated in Fig. 4. They are
characterized by the same complex fractal dimension $D_f= 
\frac{\ln{6}}{\ln{3}} +
i \frac{2\pi}{\ln{3}}$ but having different lacunarity, i.e. different measures
of the heterogeneity \cite{lacunarity}. The RSC's of Fig. 5 are 
characterized by
different fractal dimensions and connectivity, or {\it ramification} $R$. This
quantity is defined as the minimum number of bonds which should be cut in order
to remove a macroscopic part of the lattice. The RSC of Fig. 5a has an infinite
ramification, though for all others in Figs. 4-5, the ramification 
$R$ is finite.

Let each site $j$ of a RSC be allowed to contain a finite number of states or
entities $z_j = \left \{ 0,1,2,...,z_j^c \right \}$, where $z_j^c$ is hereby
taken equal to $R$.

At each step of the BTW-like dynamical process \cite{btw}, one lattice site $j$
is chosen at random and its content is updated following \begin{equation} z_j
\rightarrow z_j + 1 , \end{equation} i.e. accumulating entities on 
the site $j$.
However if $z_j \ge z_j^c$, the $j$ site is assumed to become unstable (or
``active") and to relax (in other words an avalanche is initiated) according to
the following rules \begin{equation} z_j \rightarrow z_j - z_j^c \end{equation}
\begin{equation} z_k \rightarrow z_k + 1 \end{equation} where $k$ denotes the
$z_j^c$ nearest neighboring sites of $j$. The dissipation rule is repeated $t$
times until the system reaches a stable state with all lattice sites 
$m$ implied
in the avalanche having $z_m < z_m^c$. By definition, the size $s$ of the
avalanche is the number of sites visited by the relaxation process after each
perturbation. The duration of the avalanche is $t$. Another (or the same) site
$j$ is next chosen and the (5)-(7) process repeated. One should remark that the
borders of the square lattice on which the RSC is built play the role of
absorbing sites for the dissipative process \cite{border}.

\begin{figure} \centering \includegraphics[width=.9\textwidth]{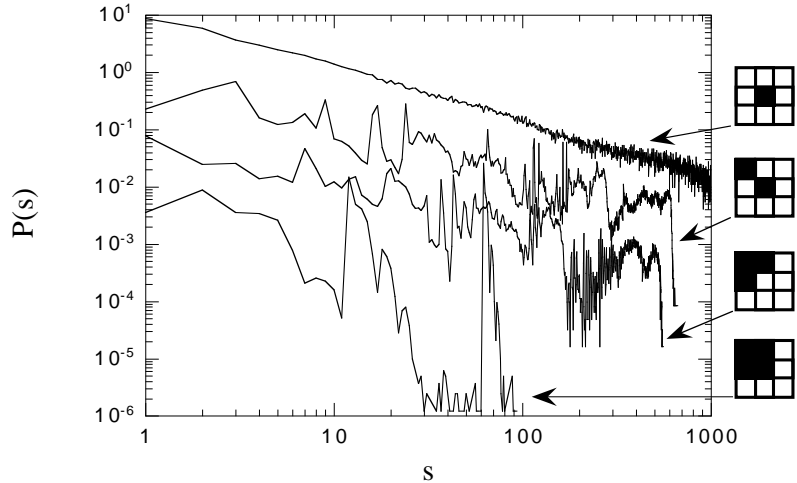}
\caption{The size distribution of avalanches $P(s)$ for the BTW model 
on fractal
RSC lattices with high and low connectivity and various  $D_f$. The 
generators of
the RSC are indicated } \label{eps5} \end{figure}

In Figs. 4-5, the distribution of avalanche size $P(s)$ is plotted for these
different RSC lattices, the generators being indicated in the margin. Since the
generators have the same size $3 \times 3$, the imaginary part of $D_f$ is
$\frac{2\pi}{\ln{3}}$ for the illustrated lattices. About $10^6$ 
avalanches have
been counted in each $P(s)$ distribution, rescaled by some arbitrary factor for
clarity. Different types of behaviors can be observed ranging from jagged
distributions with well defined peaks and valleys to ''classic'' 
smooth power law
$P(s)$ distributions. For most distributions  $P(s)$ $\sim$ $s^{-\tau} $,
expressing the scale invariance of the dynamics. We have checked the power-law
exponent ($\tau$) as a function of the fractal dimension of RSC 
lattices and have
found that $\tau$ seems to be dependent of the real part of the 
fractal dimension
$\Re \{ D_f \}$. Notice that for $\Re \{ D_f \} \rightarrow 2$,   $\tau = 1.25
\pm 0.03$.

In order to estimate $\tau$, we have filtered the jagged curve $P(s)$
distribution obtained on lattices of size $n$=3, 4, 5 and 6. As done in
\cite{gasket}, we have extrapolated the values of $\tau$ for $n \rightarrow
\infty$ in order to minimize finite-size effects. Nevertheless, error bars are
large (about 10\%) due to the presence of the oscillations. We have observed
significant deviations of $\tau$ from 1.25, i.e. the $d=2$ value. It seems that
the real part of the fractal dimension of the dissipative system is not the
single parameter controlling the value of $\tau$. Usually when the dimension of
the RSC lattice has a finite imaginary part $i \omega$, 
one\footnote{D. Stauffer
(private communication) considers that the displayed data is not convincingly
log-periodic though does not appear to be a set of random numbers.} can observe
periodic structures in $P(s)s^{\tau}$ with a period 
$\frac{2\pi}{\omega}$. At the
bottom of Fig. 5, there are huge peaks which are log-periodically spaced. These
oscillations (peaks) can be thought to originate from the DSI of the 
RSC lattice
as in \cite{gasket}, and to mimic those discussed in the main text 
for financial
indices.  We have noticed that a finite value of the ramification $R$ 
corresponds
to the largest amplitude of the oscillations. One should remark that previous
investigations \cite{gasket,luc} did not find huge oscillations nor 
sharp peaks.
The authors considered a Sierpinski gasket having loops in the 
structure as well
as a constant threshold $z_j^c=4$ everywhere on the gasket. We 
emphasize that the
connectivity of the lattice is one of the most relevant parameters. Notice that
such log-periodic oscillations and linearly substructured peaks are observed in
the time distribution of avalanche durations $P(t)$ as well.

 \end{document}